\documentclass[10pt, aps, prd, amssymb, amsmath, amsfonts, showpacs, floatfix, twocolumn, twoside, a4paper, superscriptaddress, longbibliography, nofootinbib]{revtex4-2}
\usepackage[letterpaper,top=1.45cm,bottom=1.65cm,left=1.75cm,right=1.75cm]{geometry}
\pdfoutput=1
\usepackage{verbatim}
\usepackage{comment}
\usepackage{graphicx}
\usepackage[T1]{fontenc}
\usepackage{epsfig}
\usepackage{bm}
\usepackage{tensor}
\usepackage{amssymb}
\usepackage{float}
\usepackage{amsmath}
\usepackage{subfigure}
\usepackage{dcolumn}
\usepackage[utf8]{inputenc}
\usepackage{cancel}
\usepackage[colorlinks]{hyperref}
\usepackage[usenames,dvipsnames]{color}
\usepackage{lmodern}
\usepackage[x11names]{xcolor}
\usepackage{array, colortbl, hhline}
\usepackage{xparse}
\DeclareExpandableDocumentCommand\emptycell{O{|c|}m}{\multicolumn{#2}{#1}{}}

\usepackage{fancyhdr}
\fancypagestyle{plain}

\usepackage{placeins}

\hypersetup{
     breaklinks=true,
    pdfstartview={FitH},    
    colorlinks=true,       
    linkcolor=blue,          
    citecolor=red,        
    filecolor=magenta,      
    urlcolor=blue,           
    anchorcolor=green,      
    linktocpage=true
}

\allowdisplaybreaks

\usepackage{balance}
\usepackage[normalem]{ulem}

\def\beq{\begin{equation}}
\def\eeq{\end{equation}}
\newcommand{\bea}{\begin{eqnarray}}
\newcommand{\eea}{\end{eqnarray}}

\newcommand{\mhduet}{{\texttt{MHDuet}~}} 

\definecolor{color1}{RGB}{191, 0, 255}

\begin{document}
\title{From mergers to collapse: scalarisation dynamics in neutron star binaries}
\author{Llibert Aresté Saló}
\email{llibert.arestesalo@kuleuven.be}
\affiliation{Instituut voor Theoretische Fysica, KU Leuven. Celestijnenlaan 200D, B-3001 Leuven, Belgium. }
\affiliation{Leuven Gravity Institute, KU Leuven. Celestijnenlaan 200D, B-3001 Leuven, Belgium. }

\author{Ricard Aguilera-Miret}
\email{ricard.aguilera.miret@uni-hamburg.de}
\affiliation{University of Hamburg, Hamburger Sternwarte, Gojenbergsweg 112, 21029, Hamburg, Germany}
\affiliation{Institute of Applied Computing \& Community Code (IAC3),  Universitat  de  les  Illes  Balears,  Palma, E-07122,  Spain}

\author{Miguel~Bezares}
\email{miguel.bezaresfigueroa@nottingham.ac.uk}
\affiliation{Nottingham Centre of Gravity,
Nottingham NG7 2RD, United Kingdom}
\affiliation{School of Mathematical Sciences, University of Nottingham,
University Park, Nottingham NG7 2RD, United Kingdom}

\author{Thomas~P.~Sotiriou}
\email{thomas.sotiriou@nottingham.ac.uk}
\affiliation{Nottingham Centre of Gravity,
Nottingham NG7 2RD, United Kingdom}
\affiliation{School of Mathematical Sciences, University of Nottingham,
University Park, Nottingham NG7 2RD, United Kingdom}
\affiliation{School of Physics and Astronomy, University of Nottingham,
University Park, Nottingham NG7 2RD, United Kingdom}

\begin{abstract}
We present the first fully non-linear evolutions of binary neutron star mergers in a moving-punctures approach in Einstein–scalar–Gauss–Bonnet gravity. We study both linear and quadratic-type couplings between the scalar and the Gauss-Bonnet invariant, and uncover new post-merger phenomena. These include an enhancement of the prompt collapse of a long-lived hyper-massive neutron star remnant and cases where the remnant develops a scalar configuration due to different scalarisation instabilities. This study initiates the exploration of beyond-General-Relativistic effects enhanced by the non-linear dynamics of the neutron star's fluid.
\end{abstract}

\maketitle
\thispagestyle{fancy}

      \begin{figure*}[t]
	\centering
  \includegraphics[width=0.99\linewidth, trim={14.5 0 0 0}, clip]{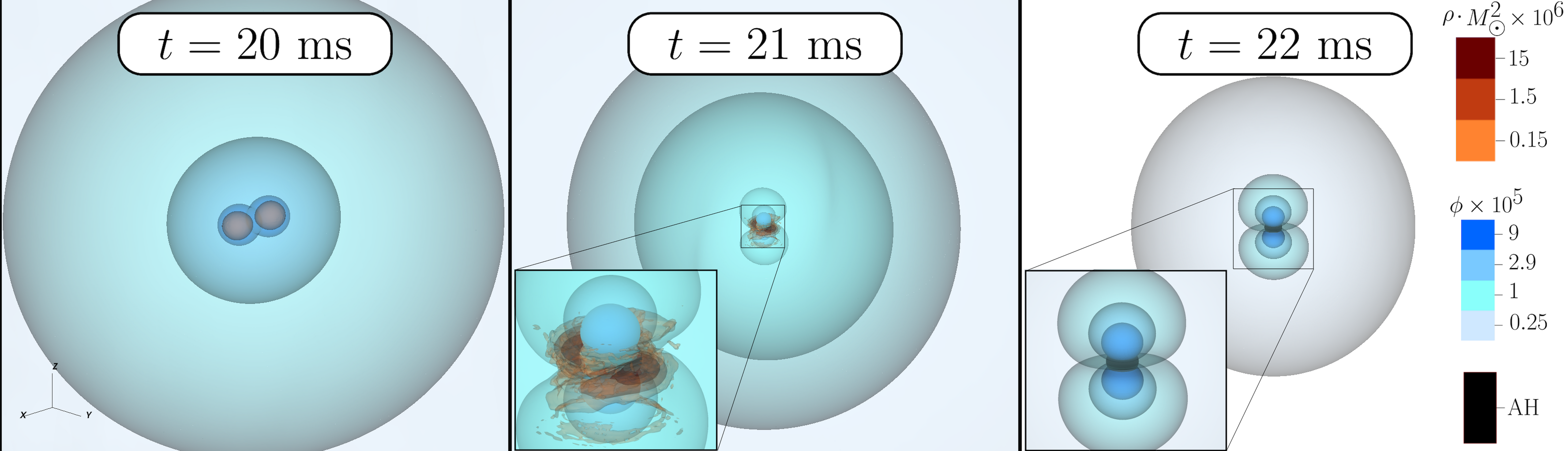}
 	\caption{\emph{Spin-induced scalarisation of a BNS merger which promptly collapses to a BH.} Snapshots of the rest-mass fluid density (shades of red) and scalar cloud (shades of blue) for a BNS merger in EsGB theory with a quadratic-type coupling of the form \eqref{eq:coupling}. The  \emph{left panel} shows the NSs during the late inspiral, which have a non-trivial scalar monopole. The \emph{centre panel} displays the prompt collapse of the remnant to a BH, when some matter is engulfed by the BH, and this causes a reconfiguration of the scalar field.
    The \emph{right panel} shows the final form of the remnant BH, with a scalar cloud that is stable due to the BH's spin. The black contour indicates the location of the apparent horizon of the BH.}
    \label{fig:3D}
    \end{figure*}

\section{Introduction}
The detection of Gravitational Waves (GWs)~\cite{LIGOScientific:2016aoc, LIGOScientific:2017vwq} has opened a new observational window into the strong-field regime of gravity. The extreme gravitational potentials and spacetime curvatures during the coalescence of compact objects, which will be accessed with increasing precision by future detectors~\cite{LISA:2024hlh,Punturo:2010zz,ET:2019dnz,ET:2025xjr,Reitze:2019iox}, offer us a promising prospect to test General Relativity (GR) and potentially observe new effects~\cite{Barack:2018yly,Barausse:2020rsu,LISA:2022kgy,Kalogera:2021bya}.

To fully exploit these detections, it is essential to obtain faithful theoretical predictions from well-motivated alternative theories of gravity. This motivates numerical simulations of compact object mergers in such scenarios. 
Attention has mostly focused on theories with an additional scalar field, as light scalars are ubiquitous in extensions of GR and the Standard Model. Several numerical studies of binary neutron star (BNS) mergers have been conducted in scalar-tensor theories of gravity within the class of theories developed by Damour and Esposito-Farèse~\cite{Barausse:2012da,Palenzuela:2013hsa,Shibata:2013pra,Kuan:2023hrh,Lam:2024wpq,Lam:2024azd} and in the presence of (kinetic) screening mechanisms~\cite{terHaar:2020xxb,Bezares:2021yek,Bezares:2021dma,Shibata:2022gec,Cayuso:2024ppe}.

Einstein–scalar–Gauss–Bonnet (EsGB) gravity, in which a scalar field couples to the Gauss-Bonnet curvature invariant, has also received a lot of attention recently, as this coupling
is known to endow black holes (BHs) with scalar hair~\cite{Kanti:1995vq,Yunes:2011we,Sotiriou:2013qea,Sotiriou:2014pfa,Sotiriou:2015pka} and trigger highly non-linear strong gravity phenomena, such as scalarisation~\cite{Damour:1993hw, Silva:2017uqg,Doneva:2017bvd,Dima:2020yac,Doneva:2022ewd}. 
Taking into account no-hair theorems \cite{Hawking:1972qk,Bekenstein:1995un,Sotiriou:2011dz,Hui:2012qt,Sotiriou:2015pka}, their known exceptions \cite{Kanti:1995vq,Yunes:2011we,Sotiriou:2013qea, Sotiriou:2014pfa, Silva:2017uqg, Doneva:2017bvd}, and field redefinitions~\cite{Weinberg:2008hq}, the scalar-Gauss-Bonnet interaction can be seen as the most relevant term of an effective field theory (EFT) of gravity with an additional scalar in the strong field regime. Hence, EsGB gravity is among the few scenarios where strong-field deviations from GR could produce observable GW signatures that could be used to search for a new light field.

Couplings between a scalar and the Gauss-Bonnet invariant can change the character of the field equations as partial differential equations, as they modify their principal part (higher derivatives). This can render the initial value problem ill-posed in conventional formulations, in which case the theory would cease to be predictive. Early simulations circumvented this problem by working perturbatively in the coupling constant, e.g.~\cite{Benkel:2016rlz,Benkel:2016kcq,Okounkova:2017yby,Witek:2018dmd, Okounkova:2019dfo}.  Such simulations, however, suffer from secular growth of errors \cite{Okounkova:2017yby} and do not capture the full non-linear dynamics beyond GR.

It has been shown that introducing a modified harmonic gauge can lead to a well-posed formulation without the need for any perturbative treatment~\cite{Kovacs:2020pns,Kovacs:2020ywu}.  The proof relies on the assumption that high-derivative contributions remain small compared to the GR terms, which is a well-motivated assumption from an EFT perspective.
This framework has subsequently been adapted to singularity-avoiding, puncture-based gauge formulations through a modified conformal and covariant Z4 approach (mCCZ4) \cite{AresteSalo:2022hua,AresteSalo:2023mmd}. These developments led to the first fully non-linear evolutions in such theories-- in particular, in EsGB-- 
for black hole spacetimes~\cite{East:2020hgw}, followed later by others~\cite{East:2021bqk,AresteSalo:2022hua,Corman:2022xqg,AresteSalo:2023mmd,Doneva:2023oww,Doneva:2024ntw,Corman:2024cdr,Lara:2025kzj,AresteSalo:2025sxc,Corman:2025wun,Shum:2025lgp,Capuano:2026lhs}. By contrast, the dynamics of neutron stars in EsGB remains still unexplored, with only two studies so far in the fully non-linear regime considering BNS~\cite{East:2022rqi} and BH-NS~\cite{Corman:2024vlk} systems.

In this work, we present the first fully non-linear simulations of BNS mergers in EsGB gravity using a moving-puncture approach. We study the shift-symmetric sector of the theory, considering both BNS systems that form a long-lived hypermassive neutron star (HMNS) remnant and systems undergoing prompt collapse to a BH. We further identify configurations that form a long-lived HMNS in GR but collapse once the Gauss--Bonnet coupling exceeds a critical strength. Finally, we investigate a quadratic-type coupling and find spontaneous scalarisation triggered either by the mass of the HMNS remnant or by the spin of the newly formed BH.

We follow the conventions in~\cite{Wald:1984rg}. Greek letters $\mu,\nu,\ldots$ denote spacetime indices. 
We set $G=c=1$.

\section{Theory}
The action of EsGB gravity reads
\begin{equation}
\label{eq:action}
    \begin{aligned}
    I = \int d^4x\sqrt{-g}&\Big[\frac{R}{16\pi}-\frac{1}{2}(\nabla\phi)^2+\lambda\,f(\phi)\,\mathcal{L}_\text{GB}\Big]+S_{\rm matter}\,,
    \end{aligned}
\end{equation}
where $\lambda$ is a coupling constant with dimensions of $[\text{length}]^2$, $S_{\rm matter}$ accounts for the neutron star's fluid, ${\mathcal L}_{\text{GB}}=R^2-4R_{\mu\nu}R^{\mu\nu}+R_{\mu\nu\rho\sigma}R^{\mu\nu\rho\sigma}$ is the Gauss-Bonnet curvature, and $f(\phi)$ is an arbitrary function of the scalar field $\phi$. We will take $f(\phi)$ to be either linear, i.e. $f(\phi)=\phi$, or quadratic-type, i.e. $f(\phi)=\phi^2/2+\mathcal{O}(\phi^3)$.

A linear coupling corresponds to a shift-symmetric theory, in which constant shifts of the scalar field preserve the action \cite{Sotiriou:2013qea,Sotiriou:2014pfa}. All compact objects in this theory are known to possess a non-trivial configuration of the scalar field, colloquially referred to as scalar hair. However, there is a notable difference between the scalar configuration of a BH and that of a NS. For BHs, the scalar exhibits the usual asymptotic fall-off, where $\phi_0$ is the scalar field value at spatial infinity and $Q_{\rm SF}$ is the scalar monopole (or scalar charge).\footnote{For clarity, we will use the term monopole, rather than charge, for the rest of the paper, to emphasise that higher multipoles are non-zero.} For NSs $Q_{\rm SF}=0$, meaning that the fall-off of the scalar field is faster than $1/r$ \cite{Yagi:2015oca}. This property of NSs in shift-symmetric EsGB gravity is essential for evading binary pulsar constraints.

A quadratic-type coupling function, namely $f(\phi)=\phi^2+\mathcal{O}(\phi^3)$, leads to some drastic differences and a richer phenomenology. Firstly, there can be both $\phi=0$ BHs and NSs stationary solutions that are identical to their GR counterparts, and scalarised solutions with $\phi\neq0$, which deviate from GR \cite{Silva:2017uqg,Doneva:2017bvd,Dima:2020yac,Herdeiro:2020wei,Doneva:2022ewd}. For the latter, both BHs and NSs have a non-zero scalar monopole (charge). For NSs, the Gauss-Bonnet invariant can take both positive and negative values within different regions of the NS, even for vanishing rotation. Therefore, scalarisation can be triggered for both signs of the coupling constant \cite{Silva:2017uqg}.
Transitions between scalarised and unscalarised configurations can occur during the inspiral-merger-ringdown signal of binary compact objects \cite{Palenzuela:2013hsa,Silva:2020omi,Elley:2022ept,Kuan:2023trn,AresteSalo:2023mmd,Capuano:2026lhs}.
The specific form of the coupling function we will use here is
\begin{equation}\label{eq:coupling}
    f(\phi)=\frac{1}{2\beta}(1-e^{-\beta\phi^2})\,,
\end{equation}
where $\beta$ is a dimensionless parameter. This choice improves stability compared to the simpler $f(\phi)=\phi^2$ coupling~\cite{Blazquez-Salcedo:2018jnn,Silva:2018qhn,Macedo:2019sem}.

\begin{figure*}[t]
	\centering
    \includegraphics[width=0.49\linewidth]{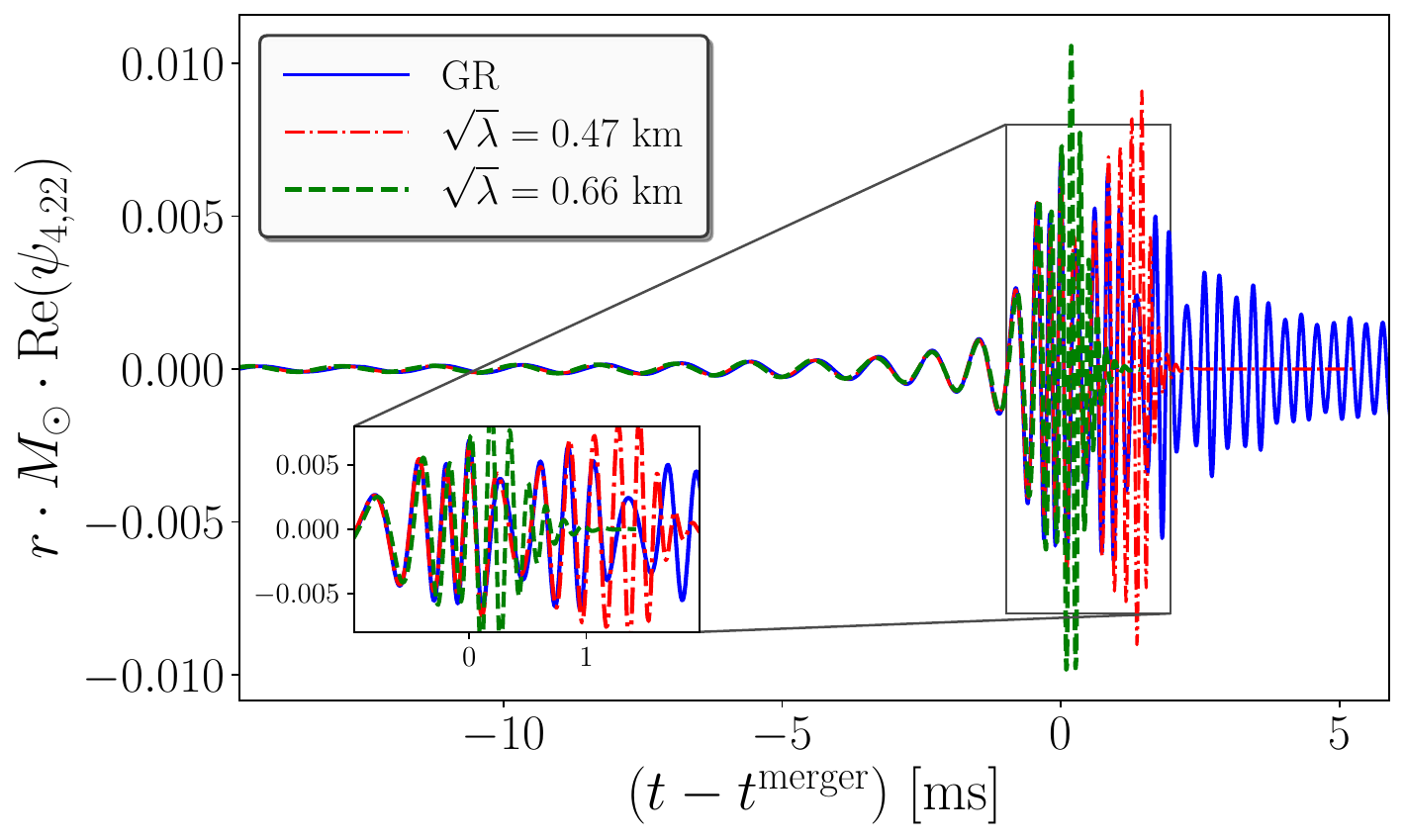}
	\includegraphics[width=0.49\linewidth]{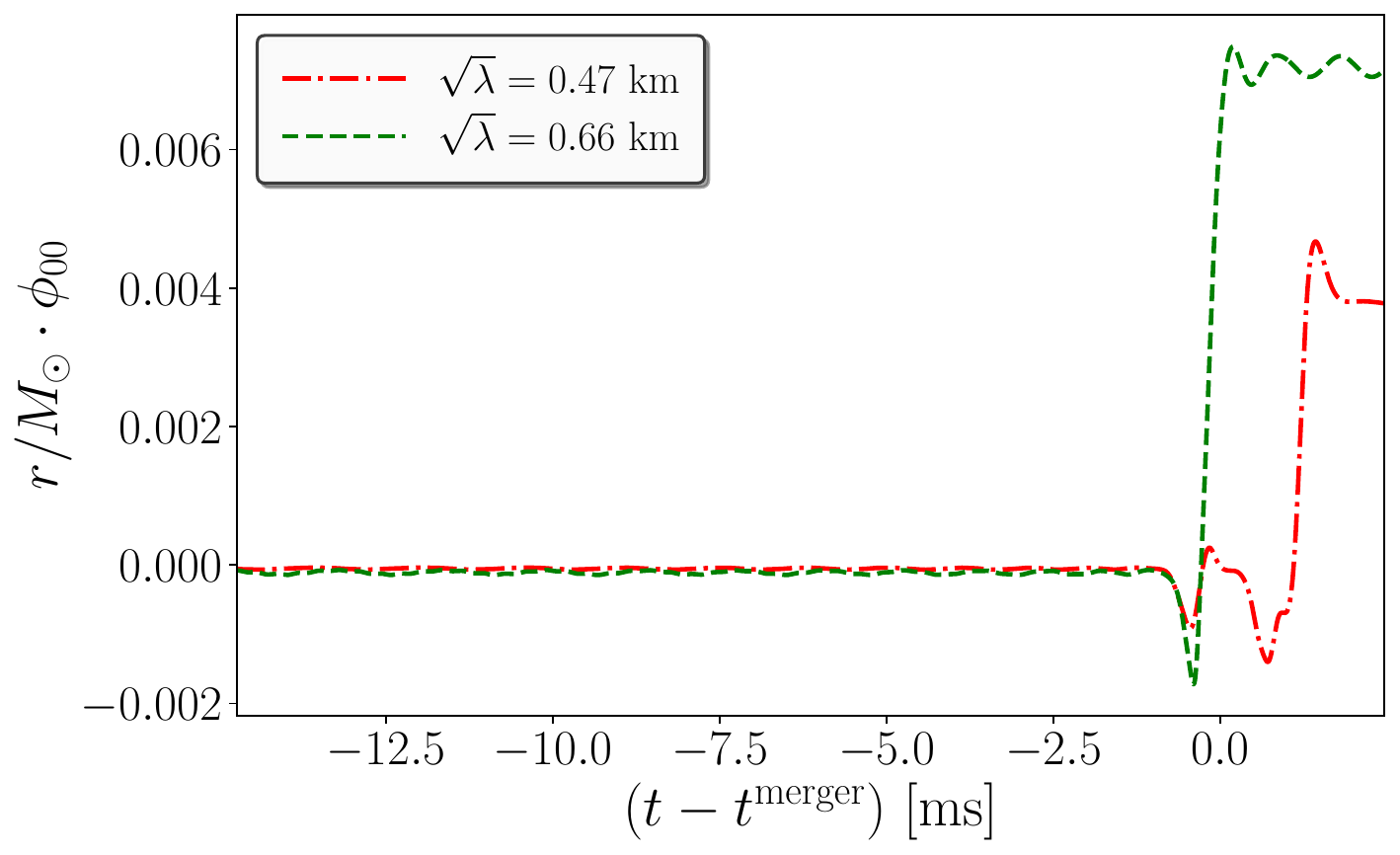}
 	\caption{\emph{Prompt collapse of a BNS merger to a BH enhanced by shift-symmetric EsGB gravity.} The left panel shows the $(2,2)$ mode of the Newman-Penrose $\Psi_4$ scalar of the GW, while the right panel displays the $(0,0)$ mode of the scalar field, both extracted at $r=100M_{\odot}$. We consider two values of the coupling constant: for the weaker one, the collapse occurs after the formation of the HMNS remnant, whereas for the stronger one, it occurs as soon as the merger occurs. The scalar monopole only becomes non-zero after the appearance of the BH.}
    \label{fig:scalar}
    \end{figure*}

\

\section{Initial data and set-up.}
All simulations are initialised from constraint-satisfying general-relativistic initial data for the metric and fluid variables, constructed using the {\sc Lorene}~\citep{lorene} package. The scalar field is then consistently superposed on these data, and the full EsGB system is then evolved dynamically. For shift-symmetric EsGB gravity, we initialise the scalar field by setting $\phi=\partial_t\phi=0$, which constitutes a valid solution of the shift-symmetric theory at the level of the initial data. During the subsequent evolution, the system experiences a short initial relaxation phase in which the neutron stars dynamically acquire a non-trivial configuration of the scalar field. This transient can slightly affect the orbital dynamics of the inspiral. Unless otherwise stated, this initial relaxation phase is not shown in the figures. For quadratic-type couplings, we instead initialise the scalar field with $\partial_t\phi=0$ and a small Gaussian seed $\phi=A\exp[-(r-r_0)^2/(2\sigma^2)]$, with $\sigma=1$, $A=10^{-5}$, and $r_0=90\,M_{\odot}$, in order to trigger the scalarisation instability. The resulting constraint violations are mild and efficiently damped during the initial transient by the CCZ4 formulation.

We consider both isolated NSs and BNS systems. Single NSs are placed at the centre of the numerical domain and evolved for several values of the mass and angular momentum. Binary systems are initialised at coordinate separations of either $45$ or $50$~km, depending on the total mass, allowing for several orbits prior to merger. We focus on equal-mass BNS configurations with ADM masses $M_{\rm ADM}=\{2.7,\,2.6,\,3.0\}M_{\odot}$, such that the first two cases do not undergo prompt collapse in GR, while the last case does. The corresponding initial orbital angular frequencies are $\Omega=\{1774.68,\,1738.42,\,1597.35\}\,\mathrm{rad\,s^{-1}}$, as summarised in Table~\ref{tab:sim_params}. In all simulations, the neutron stars are modelled using a tabulated piecewise-polytropic representation of the zero-temperature APR4 equation of state.  Finally, constructing fully constraint-satisfying initial data for generic scalar-field configurations in EsGB gravity remains an open problem, although promising progress has recently been made~\cite{Brady:2023dgu,Nee:2024bur}. Further details on the formulation, numerical implementation, and code validation are provided in Appendix~\ref{sec:app_formulation}.

\begin{table}[htbp]
      \centering
      \setlength{\tabcolsep}{8pt}
      \caption{\emph{Parameters of the BNS simulations presented}}
      \begin{tabular}{|c|c|c|c|c|c|}
        \hline
        & \big($m_1$, $m_2$\big)  & Initial sepa-  & Initial angular\\
        & $[M_{\odot}]$ & ration [km] & frequency [rad s$^{-1}$]\\
        \hline
        I & (1.35, 1.35) & 45 & 1774.68 \\\hline           
         II & (1.3, 1.3) & 45 & 1738.42 \\   \hline
         III & (1.5, 1.5) & 50 & 1597.35\\
        \hline
      \end{tabular}
      \label{tab:sim_params}
    \end{table}

The coupling constant values used for shift-symmetric EsGB BNS systems are $\lambda=\{0.1,0.2\}M_{\odot}^2$ corresponding to $\sqrt{\lambda}=\{0.47,0.66\}$ km. These values are slightly larger than the current most stringent observational bounds on the theory, which come from comparing the black hole-neutron star GW signal GW230529 to Post-Newtonian (PN) results for EsGB, yielding a constraint of $\sqrt{\lambda}\lesssim 0.28$ km \cite{Sanger:2024axs}. 
For quadratic-type couplings, we instead use $\lambda=\{150,-90\}M_{\odot}^2$ corresponding to $\sqrt{|\lambda|}=\{18.05,13.98\}$ km, which we can compare to the current observational limits for pulsars studied in \cite{Danchev:2021tew}. For the negative value of the coupling, the corresponding constraint for the APR4 EoS yields $\sqrt{|\lambda|}\lesssim 7.36$ km, which is only slightly below the value considered. Instead, for the positive value of the coupling, the constraints were only derived for other EoS, with bounds ranging from $11.43$ to $26.05$ km. Nevertheless, they all remain within the same order of magnitude and, in addition, slightly higher values are also allowed, as pointed out in \cite{Wong:2022wni}, where it was shown that strongly disfavoured values only start at $\sqrt{\lambda}\gtrsim 41.27$ km.


\

\begin{figure*}[t]
	\centering
	\includegraphics[width=0.49\linewidth]{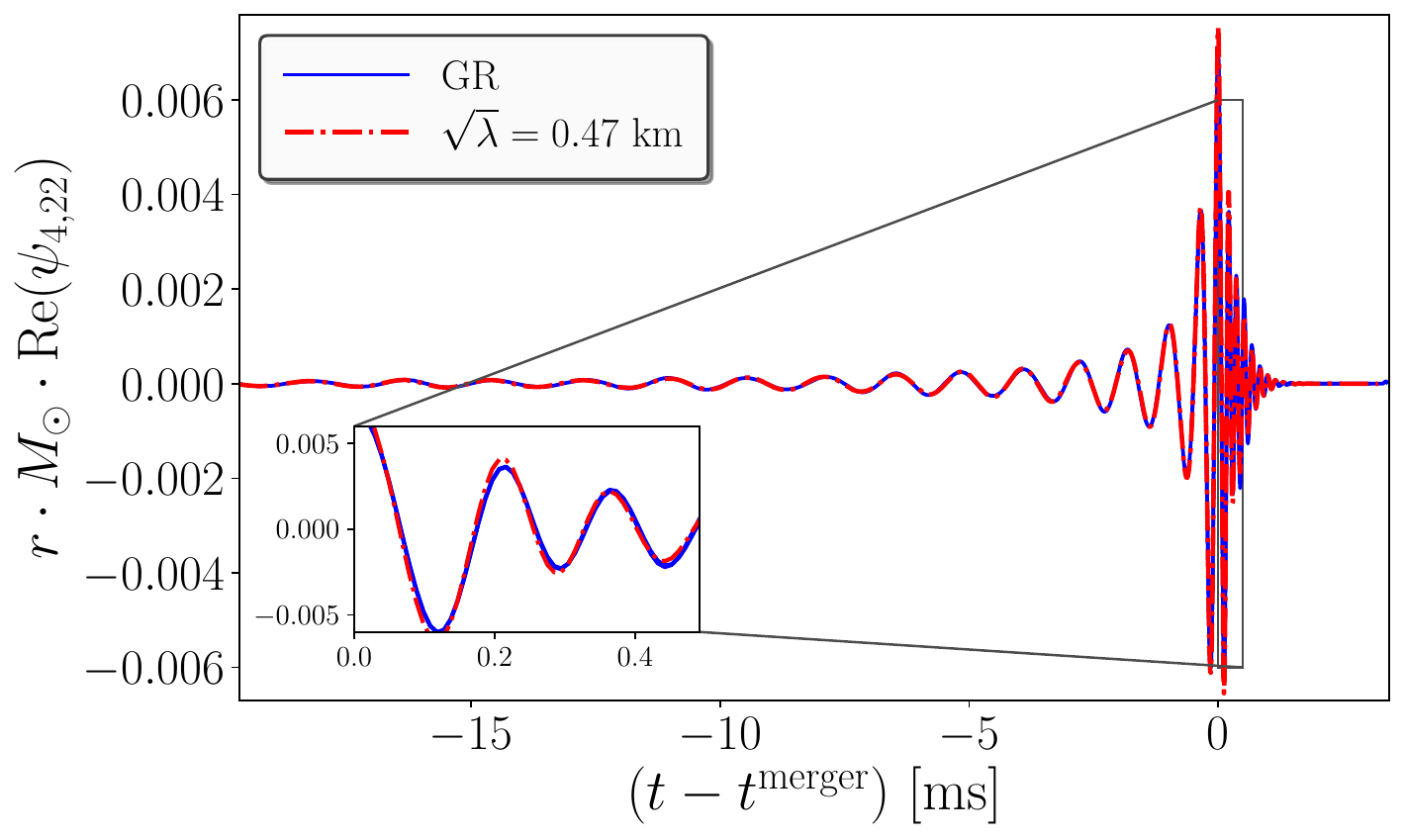}
    \includegraphics[width=0.49\linewidth]{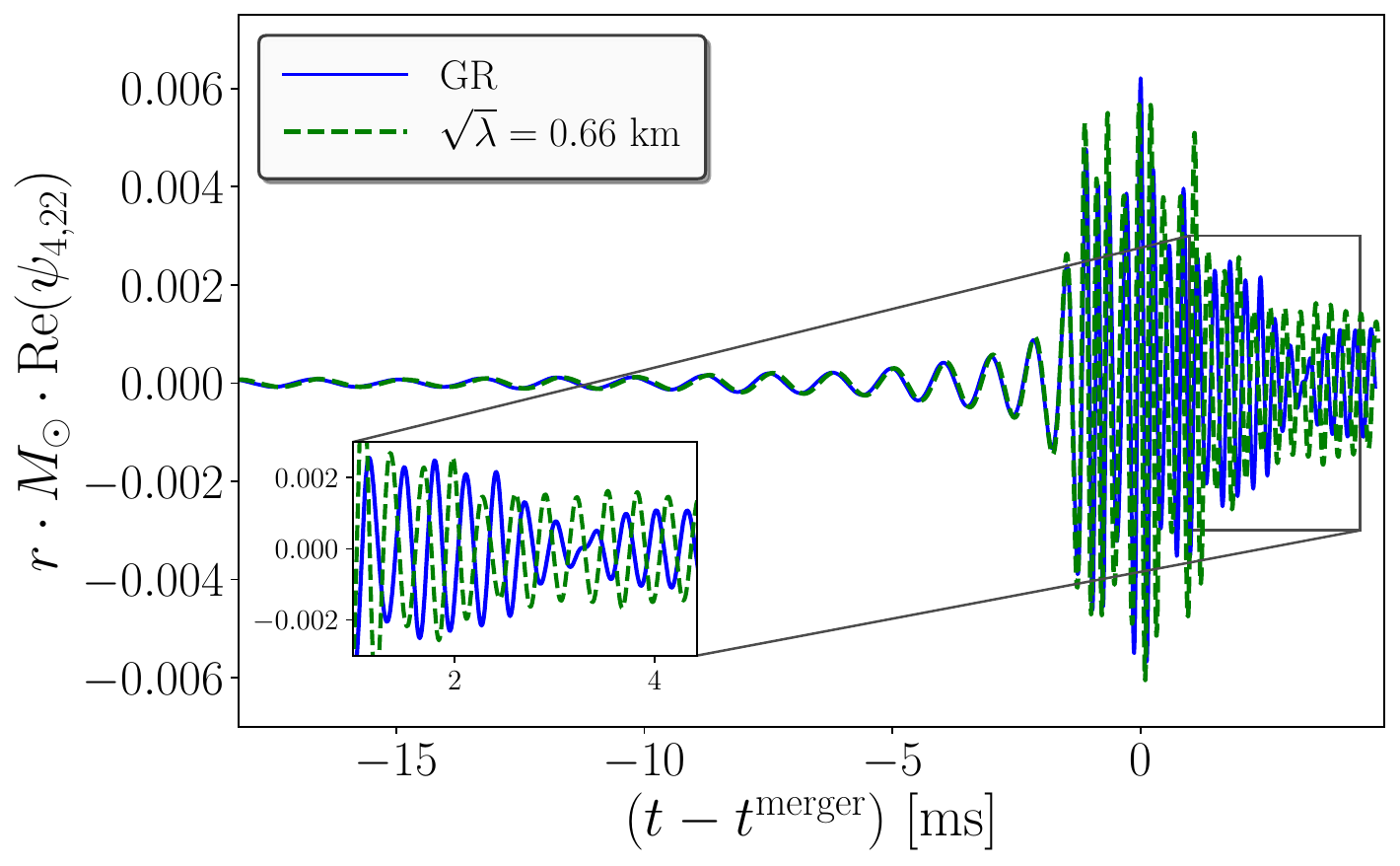}
	\caption{Here we show the gravitational waveforms for BNS systems, both in GR and beyond, with higher (\emph{left}) and lower (\emph{right}) total mass with respect to the system in Fig.~\ref{fig:scalar}. Both GR and EsGB systems on the left promptly collapse into a BH with no visible dephasing, likely due to the lack of scalar monopole of the NSs. Both systems on the right lead to a long-lived HMNS remnant with some visible differences in the post-merger signal.}
    \label{fig:coll_and_low}
    \end{figure*}

\section{Shift-symmetric EsGB binary neutron stars} 
We consider BNS mergers in shift-symmetric EsGB for the three types of initial data shown in Table \ref{tab:sim_params}. In the left panel of Fig.~\ref{fig:scalar}, we show the GW signal of a BNS system that leads to a long-lived HMNS remnant in GR but collapses into a BH in EsGB once the coupling constant exceeds a critical value, corresponding to case I of Table~\ref{tab:sim_params}. This could be a smoking-gun signature of beyond GR physics. We note that this initial data is chosen to lie close to the prompt collapse threshold.\footnote{For an equal mass, irrotational and quasi-circular orbit using the APR4 EoS, the threshold for a prompt collapse lies at $M\simeq1.37M_{\odot}$, according to internal tests.} As a result, the additional scalar degree of freedom efficiently extracts angular momentum from the system, destabilising the HMNS remnant. Our results therefore suggest that the scalar coupling can modify the threshold for prompt collapse, although a precise determination of the magnitude of this shift will require further numerical investigation.

As we commented on earlier, NSs in shift-symmetric theory ---in contrast to BHs--- have no scalar monopole. Therefore, the dephasing of the gravitational wave with respect to GR is expected to be negligible. This is consistent with the left panel of Fig.~\ref{fig:scalar}, where waveforms have been aligned at the peak of the merger. However, longer waveforms with a more controlled eccentricity and higher resolution would be needed to perform a robust alignment and find out whether the effect recently found in \cite{Corman:2025wun} can also hold in BNS systems.

In the right panel of Fig.~\ref{fig:scalar}, we show the behaviour of the $(0,0)$ mode of the scalar waves for the EsGB simulations corresponding to case I of Table~\ref{tab:sim_params}, which collapse into a BH. It clearly shows how the inspiral NSs have no scalar monopole, while this ramps up as soon as a BH is formed. We note that the small oscillations observed during the inspiral are very likely caused by the artificial initial transient going from zero to a non-zero scalar field. This  induces oscillations in the fundamental fluid mode of the star (f-mode) \cite{East:2022rqi}, which backreacts on the monopolar charge of the scalar field. 

Fig.~\ref{fig:coll_and_low} displays cases for a lower and higher total mass, corresponding to cases II and III of Table~\ref{tab:sim_params}, respectively. The low-mass configuration forms a long-lived HMNS remnant in both GR and non-GR, whereas the high-mass undergoes prompt collapse for both cases. These results are perfectly consistent with the ones in \cite{East:2022rqi}. We further note that similar differences in the post-merger signal to those shown in the right panel are also observed for different EoS, see, e.g., ~\cite {Llorens-Monteagudo:2025mxr}. Quantifying these differences relative to numerical uncertainties, particularly in the GW phase, is left for future work. It is worth stressing that, although there is no scalar monopole for NSs in shift-symmetric EsGB \cite{Yagi:2015oca}, there is some non-trivial configuration of the scalar field which decays faster than $1/r$. This scalar field has to reconfigure itself in the moment of the merger and this has a non-linear effect which impacts the post-merger itself in a similar way as a change in the EoS.

Finally, we have explored the behaviour of the BNS systems for slightly higher coupling values. We have seen that for coupling constants $\sqrt{\lambda}=0.95$ km and $1.34$ km, we can evolve consistently throughout all of the inspiral, but we find a loss of hyperbolicity at the merger, both for systems collapsing to a BH and for those with a long-lived HMNS remnant NS. These values do coincide with the threshold of hyperbolicity observed for isolated black holes when normalised by the system's total mass. However, since the hyperbolicity bounds for BBH are dominated by the smaller object, this means that the threshold for $\lambda/(m_1+m_2)^2$ in shift-symmetric EsGB BNS systems is $(1+q)^2$ times larger than in BBH systems, with $q=m_2/m_1>1$ being the unequal mass ratio.\footnote{Using the results of~\cite{Doneva:2023oww}, we assess loss of hyperbolicity of the system with the breakdown of the simulation, which coincides with the threshold of hyperbolicity observed for isolated BHs, strongly suggesting that the system also loses hyperbolicity here when the simulations breaks down.}

\

\section{Scalarisation}
\begin{figure}[t]
	\centering
	  \includegraphics[width=1.0\linewidth]{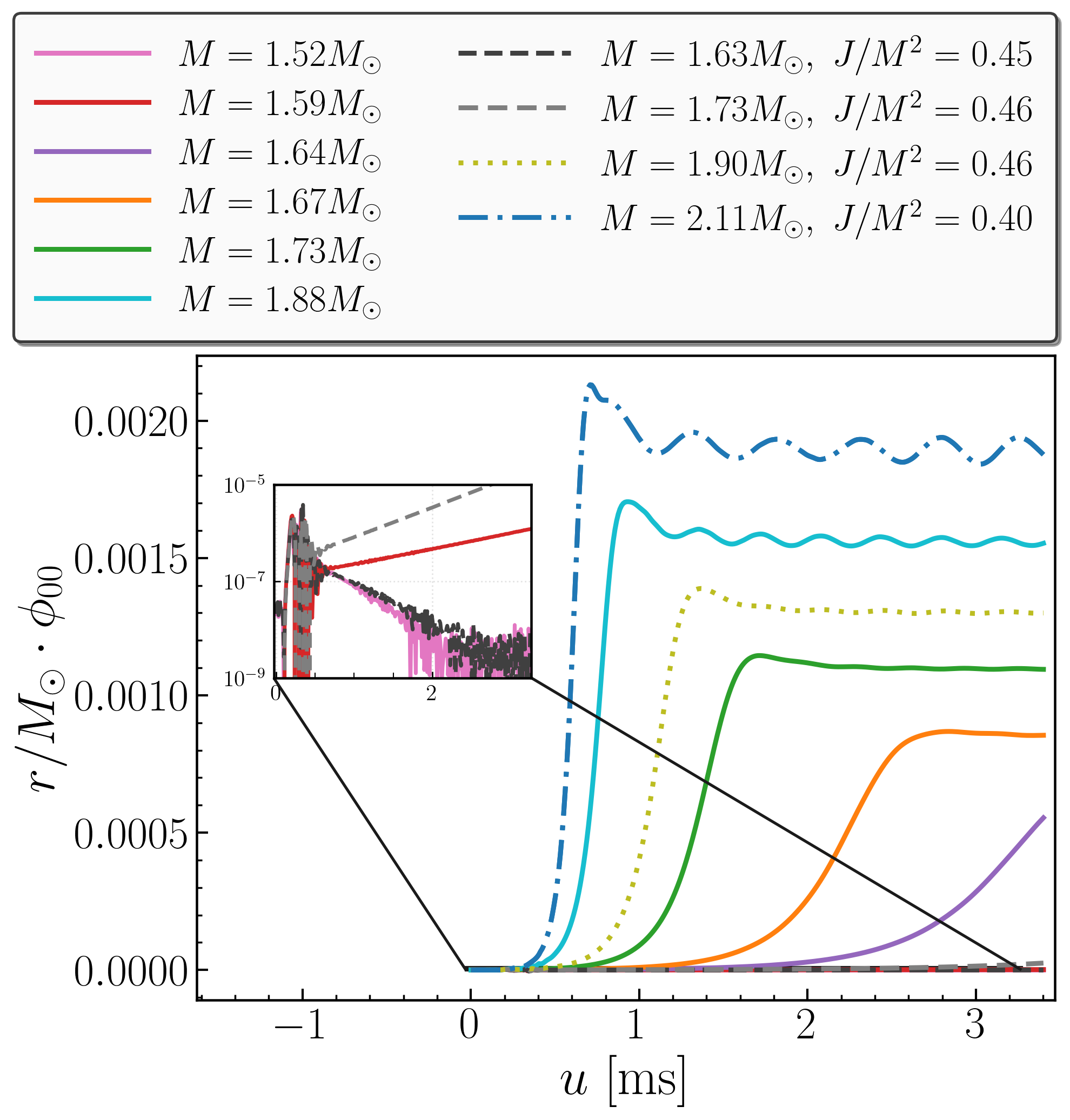}
	\caption{\emph{Spontaneous scalarisation of single NSs in EsGB with a quadratic-type coupling.} We have considered several values of neutron star's masses and spins with the coupling constants of $\lambda=150M^2_{\odot}$ and $\beta=400\cdot16\pi$. This plot shows that the threshold mass of scalarisation increases significantly for a large spin. See Appendix~\ref{sec:app_convergence} for convergence of the spinning NS with higher mass.}
    \label{fig:scalarization_ns}
    \end{figure}

We next consider quadratic-type couplings of the form of \eqref{eq:coupling}, which can lead to the phenomenon of ``spontaneous scalarisation''~\cite{Doneva:2022ewd}. As discussed above, NSs can undergo scalarisation for both signs of the coupling $\lambda$. Moreover, the scalarisation threshold depends on the spin as well. We start by investigating the scalarisation of isolated NSs for positive $\lambda$, as this will be useful for interpreting other results. The development of a scalar monopole over retarded time $u\equiv t-r_{\rm ex}-2M_{\odot}\log(r_{\rm ex}/(2M_{\odot})-1)$ (with $r_{\rm ex}=100M_{\odot}$ the extraction radius) is shown in Fig.~\ref{fig:scalarization_ns}. For non-spinning NSs configurations, it develops faster and reaches higher values for more massive NSs, in agreement with \cite{Doneva:2023kkz}. For the value of the coupling considered here, the theoretical threshold mass is $M=1.57M_{\odot}$,\footnote{We thank Nicola Franchini for having computed this value with his static spherically symmetric code.} and this is consistent with Fig.~\ref{fig:scalarization_ns}, where an NS of mass $M=1.59M_{\odot}$ experiences scalarisation, while one of $M=1.52M_{\odot}$ does not.
We next consider spinning NSs with a fixed dimensionless spin, $J/M^2\approx0.45$, and examine several mass values. We find  that the threshold mass for scalarisation is higher for spinning NSs, lying between $1.63M_{\odot}$ and $1.73M_{\odot}$. This behaviour is expected for $\lambda >0$, as rotation reduces the central
fluid energy density, thereby weakening the scalarisation
instability.

\begin{figure*}[t]
	\centering
	\includegraphics[width=0.49\linewidth]{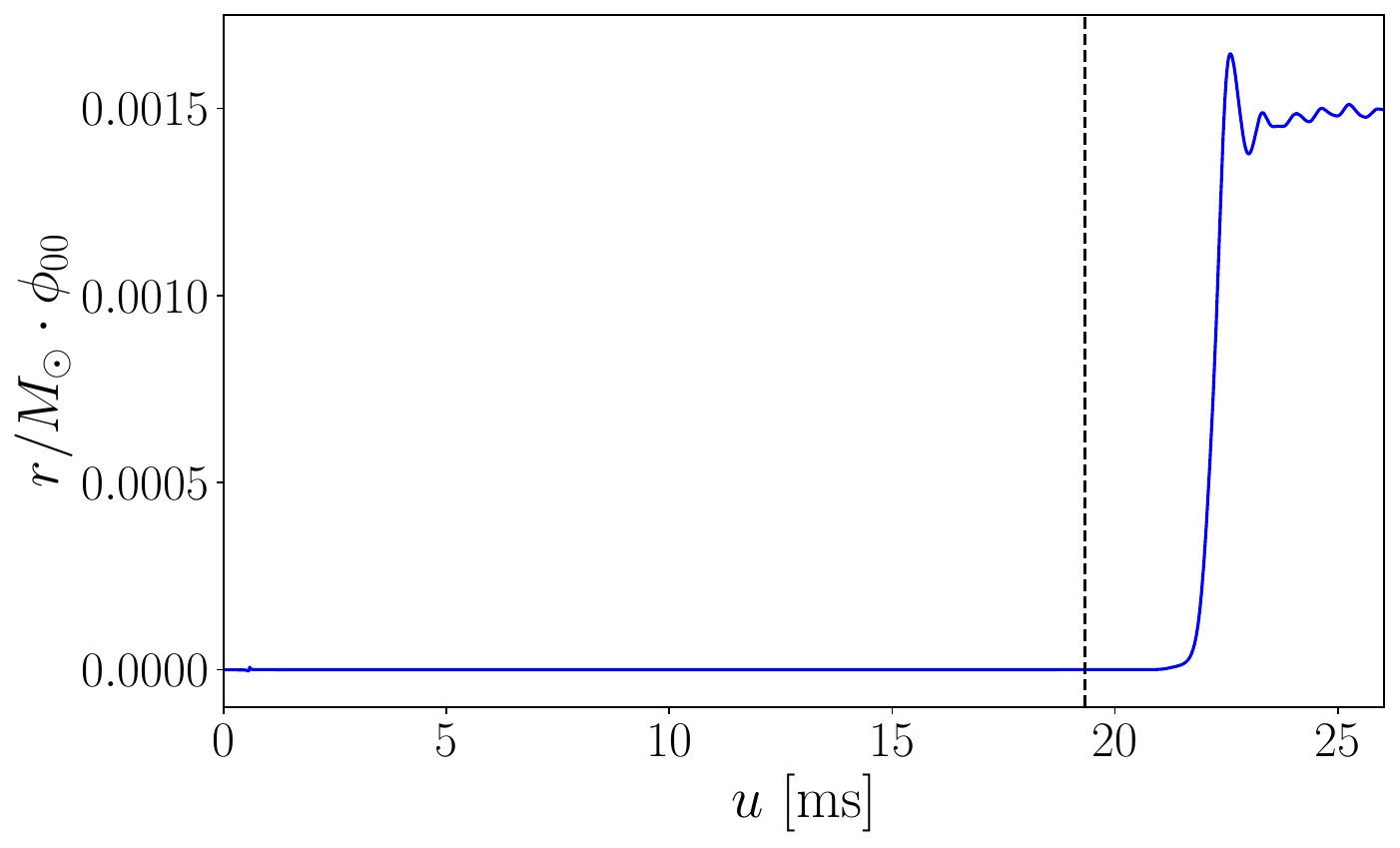}
    \includegraphics[width=0.49\linewidth]{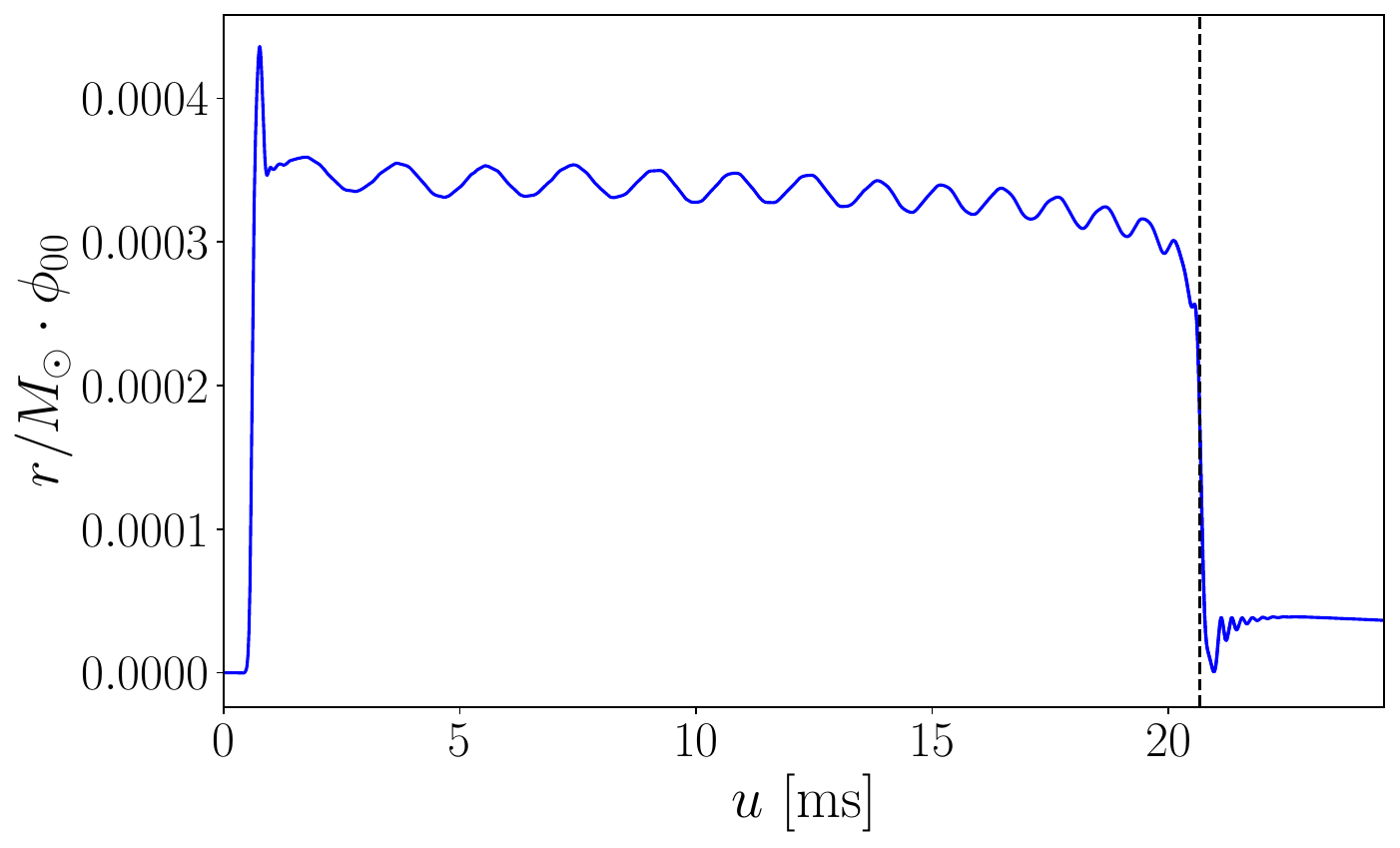}
	\caption{\emph{BNS merger in EsGB gravity with quadratic-type coupling.} The left panel shows the evolution of the scalar monopole for a BNS merger leading to a long-lived HMNS remnant in which the scalar field only develops after merger because of the larger mass of the remnant for the coupling values of $\lambda=150M^2_{\odot}$ and $\beta=400\cdot16\pi$. The right panel considers instead the case of a BNS merger promptly collapsing to a BH shown in Fig.~\ref{fig:3D}, for the coupling values of $\lambda=-90M^2_{\odot}$ and $\beta=2000\cdot16\pi$. The dotted lines indicate the time of the merger.}
    \label{fig:spont_bns}
    \end{figure*}

We now turn our attention to BNS mergers. In the left panel of Fig.~\ref{fig:spont_bns}, we show the evolution of the scalar monopole for a BNS system whose merger leads to a long-lived HMNS remnant, with $\lambda>0$. While the scalar monopole remains zero during the inspiral, scalarisation takes place post-merger. 
The HMNS formed by the merger is more massive than the inspiral NSs, so its fluid energy density is large enough to trigger the instability responsible for scalarisation. Hence, we attribute the scalarisation of the increase in mass of the NSs remnant, rather than to the spin-induced scalarisation found for BBH systems \cite{Dima:2020yac,Herdeiro:2020wei,Berti:2020kgk,AresteSalo:2023mmd}. Our simulations verify predictions obtained using perturbation theory and stationary solutions of NSs in EsGB gravity with quadratic-type couplings~\cite{Doneva:2023kkz}. Furthermore, we have included a convergence test for this run in Appendix~\ref{sec:app_convergence}.

Finally, we consider the opposite sign of the coupling constant $\lambda$ and evolve a BNS that promptly collapses into a BH. We choose such values for $\lambda$ and for the NS masses so that the BH remnant would have scalar hair due to spin--induced scalarisation. Indeed, our simulations capture this phenomenon for the first time.
The scalar monopole is shown in the right panel of Fig.~\ref{fig:spont_bns}.  For the NS masses and value of $\lambda$ we have chosen, the inspiral NSs are scalarised. The scalar monopole exhibits oscillations during the inspiral due to the initial transient and drops sharply at prompt collapse as the fluid energy density, which supports the NSs scalar configuration, is lost. 
Then, the scalar monopole grows again as a result of a spin-induced scalarisation of the BH. Several time snapshots of the three-dimensional configuration of the scalar cloud around this system are shown in Fig.~\ref{fig:3D}. Near the remnant BH, the scalar takes larger values around the poles of the apparent horizon, as these are regions where the Gauss-Bonnet invariant is negative and the scalarisation instability is hence stronger. This is consistent with the results of \cite{AresteSalo:2023mmd}.

\

\section{Conclusions}
We present the first fully non-linear binary neutron star merger evolutions using a moving-punctures approach in Einstein–scalar–Gauss–Bonnet gravity. For this purpose, we have implemented the modified CCZ4 formulation \cite{AresteSalo:2022hua} in the \mhduet code~\cite{Palenzuela:2025ucx}, demonstrating that such systems
can be evolved robustly in the fully non-linear regime. These simulations include additional physical scenarios with respect to earlier  studies~\cite{East:2022rqi} and reveal several new signatures of beyond GR physics in the GW signals of BNS mergers. Moreover, our simulations are a  demonstration of a successful implementation of the moving-punctures approach --- a very extensively used method in the numerical relativity community --- to BNS systems in EsGB gravity. Using moving punctures can be advantageous, as it eliminates the need for excision. More broadly,  being able to solve the same problem with independent codes and different numerical methods is essential for ensuring the generation of more accurate waveforms and revealing subtle phenomena (see e.g.~\cite{Lara:2025kzj,Corman:2025wun}).

We have studied both systems that lead to a long-lived HMNS remnant and to a prompt collapse of the remnant to a BH, for both linear and quadratic-type couplings. For linear couplings, NSs do not have a scalar monopole, leaving the inspiral phase effectively indistinguishable from 
GR even for significantly large values of the coupling constant. Nevertheless, our simulations show that smoking-gun signatures can arise during the merger phase, such as the prompt collapse to a BH for systems that would otherwise form a long-lived HMNS in GR. A systematic exploration of the dependence of the post-merger signal on the NS equation of state is left for future
work.

Quadratic-type couplings instead give rise to a much richer phenomenology, including nonlinear phenomena associated with scalarisation \cite{Doneva:2022ewd}. Our simulations captured for the first time two such phenomena. The first one is the scalarisation of the long-lived HMNS remnant after the merger of two unscalarised NSs. The second is the prompt descalarisation after the merger of scalarised NSs, followed by a spin-induced scalarisation of a remnant BH.

This work provides a basis for studying further interesting signatures of beyond GR effects enhanced by the dynamics of the neutron star's fluid, which can only be captured in the fully non-linear regime and complements the effects recently seen in~\cite{Lara:2025kzj,Corman:2025wun} in vacuum spacetimes. Further work is required to improve the initial data and eliminate the transient effect in the fluid's f-mode, which is especially important in cases where the NSs have scalar monopole. In future work, we plan to complete the picture by studying BH-NS systems and incorporating the effect of magnetic fields.

\

\noindent
\section*{Acknowledgments} 
We thank Katy Clough, Nicola Franchini, Ludovico Machet, Borja Miñano and Carlos Palenzuela for useful discussions. In the early stages of this work LAS was supported by a London Mathematical Society (LMS) Early Career Fellowship. LAS is partly funded by Interuniversitaire Bijzonder Onderzoeksfonds (IBOF)/21/084. RA-M is funded by the Deutsche Forschungsgemeinschaft (DFG, German Research Foundation) under the Germany Excellence Strategy - EXC 2121 `Quantum Universe' - 390833306 and by the European Research Council (ERC) Advanced Grant INSPIRATION under the European Union’s Horizon 2020 research and innovation program (grant agreement no. 101053985). MB acknowledges partial support from the STFC Consolidated Grant nos. ST/Z000424/1 and UKRI2492. TPS acknowledges partial support from the STFC Consolidated Grant nos. ST/V005596/1,  ST/X000672/1, and UKRI2492.  This work used the DiRAC Memory Intensive service Cosma8 at Durham University, managed by the Institute for Computational Cosmology on behalf of the STFC DiRAC HPC Facility (\url{www.dirac.ac.uk}). The DiRAC service at Durham was funded by BEIS, UKRI and STFC capital funding, Durham University and STFC operations grants. DiRAC is part of the UKRI Digital Research Infrastructure. We used the resources provided by the VSC (Flemish Supercomputer Center), funded by the Research Foundation - Flanders (FWO) and the Flemish Government. We acknowledge the computing time made available
to us on the high-performance computer ``Lise'' at the NHR Centre NHR@ZIB. This center is jointly supported by the Federal Ministry of Education and Research and the state governments participating in the NHR.

\appendix

\section{Formulation and implementation}\label{sec:app_formulation}
To evolve the fully non-linear EsGB equations of motion, we employ the modified conformal and covariant Z4 (mCCZ4) formalism \cite{AresteSalo:2022hua}, first implemented in \texttt{GRFolres} \cite{AresteSalo:2023hcp} (an extension of the \texttt{GRChombo} \cite{Andrade:2021rbd} numerical relativity code), which was shown to be well-posed in EsGB in the weakly coupling regime. In the same spirit as the modified harmonic gauge \cite{Kovacs:2020pns,Kovacs:2020ywu}, 
this formalism introduces two auxiliary metrics to ensure that gauge modes propagate with speeds distinct from the physical ones, defined as
\begin{equation}
    \tilde{g}^{\mu\nu}=g^{\mu\nu}-a(x)n^{\mu}n^{\nu}\,,\qquad \hat{g}^{\mu\nu}=g^{\mu\nu}-b(x)n^{\mu}n^{\nu}\,,
\end{equation}
where $a(x)$ and $b(x)$ are arbitrary functions such that $0<a(x)<b(x)$, and $n^{\mu}$ is the unit timelike vector normal to $t=\text{const}$. These choices modify the equations of motion both for the metric and gauge variables (see \cite{AresteSalo:2022hua, AresteSalo:2023mmd} for further details); throughout this work we fix $a(x)=0.2$ and $b(x)=0.4$, following previous studies.

The EsGB equations in the mCCZ4 formulation are coupled to the general relativistic magnetohydrodynamic equations and evolved using the open-source code {\texttt{MHDuet}} ~\cite{palenzuela18,Palenzuela:2025ucx,mhduet_web}, an evolution code for general relativistic magnetohydrodynamics with neutrino transport. The code is generated by the {\sc Simflowny} platform ~\cite{arbona13,arbona18} from a high-level specification of the computational system, and can target either the {\sc SAMRAI} infrastructure~\citep{hornung02,gunney16} or, in newer versions, the {\sc AMReX} framework~\cite{AMReX_JOSS, AMREXOVERVIEW, Zhang_2021} providing parallelisation and adaptive mesh refinement (AMR). The code employs a fourth-order-accurate finite-difference operator for the spatial derivatives in the metric and scalar field sectors (supplemented with sixth-order Kreiss--Oliger dissipation), a high-resolution shock-capturing (HRSC) scheme for the fluid based on Lax--Friedrichs flux
splitting~\cite{shu98} and fifth-order MP5 reconstruction~\cite{suresh97}, and a fourth-order Runge--Kutta time integrator with time step $\Delta t \leq 0.4\,\Delta x$. Refinement boundaries are treated efficiently and accurately when sub-cycling in time~\cite{McCorquodale:2011,Mongwane:2015}. A new version of {\texttt{MHDuet}} including the EsGB model, based on the {\sc AMReX} infrastructure and supporting GPU acceleration, will be made open source and released in the near future in~\cite{mhduet_web}.

The binary is evolved in a cubic domain of size $\left[-1228,1228\right]$~km in each direction. The inspiral is fully covered by six Fixed Mesh Refinement (FMR) levels. Each consists of a cube with twice the resolution of the next larger one. In addition, we use a single AMR level, covering regions where the density exceeds $5 \times 10^{12}~\rm{g~cm^{-3}}$ and providing a uniform resolution throughout the shear layer. When the HMNS remnant collapses into a BH, we instead cover the regions where the lapse is below $0.65$. With this grid structure, we achieve a maximum resolution of $\Delta x_{\rm min} = 120$ m covering at least the most dense region of the remnant.

In all simulations, we employ a tabulated version of the EoS at zero temperature which is augmented by an approximate thermal contribution to pressure and internal energy. While the largest contribution to the pressure is provided by the cold EoS, in a BNS merger the temperature cannot be neglected. By assuming an ideal gas, we can calculate the thermal contribution to the pressure as $\rho_{\rm th}=(\Gamma_{\rm th}-1)\rho\epsilon$, where $\Gamma_{\rm th}=1.8$ is the adiabatic index. This approach is frequently used in numerical relativity simulations, e.g.~\cite{Kiuchi:2017zzg,Palenzuela:2021gdo,Aguilera-Miret:2024cor}.

\

\section{Convergence test} \label{sec:app_convergence}

In this section, we study convergence for the simulations carried out in this work with a positive quadratic-type coupling yielding spontaneous scalarisation of NSs. In particular, we perform the analysis for both a single NS and a BNS merger.

\

{\em \textbf{Single NS.}---}
First, we analyse the $(0,0)$ mode of the scalar field within the spontaneous scalarisation of a single NS shown in Fig.~\ref{fig:scalarization_ns}. In particular, we study the case with $M=2.11M_{\odot}$ and $J/M^2=0.40$, which is close to the values attained by the HMNS remnant in a BNS merger system. We consider three different resolutions, namely low (LR), medium (MR) and high (HR), corresponding to maximum resolutions of $\Delta x_{\rm min} = 240,~120$ and $60$ m, respectively. The MR run uses the same set-up explained in the previous section, whereas one FMR level was added (subtracted) for the HR (LR) case.

\begin{figure}[h!]
	\centering
	 \includegraphics[width=1.0\linewidth]{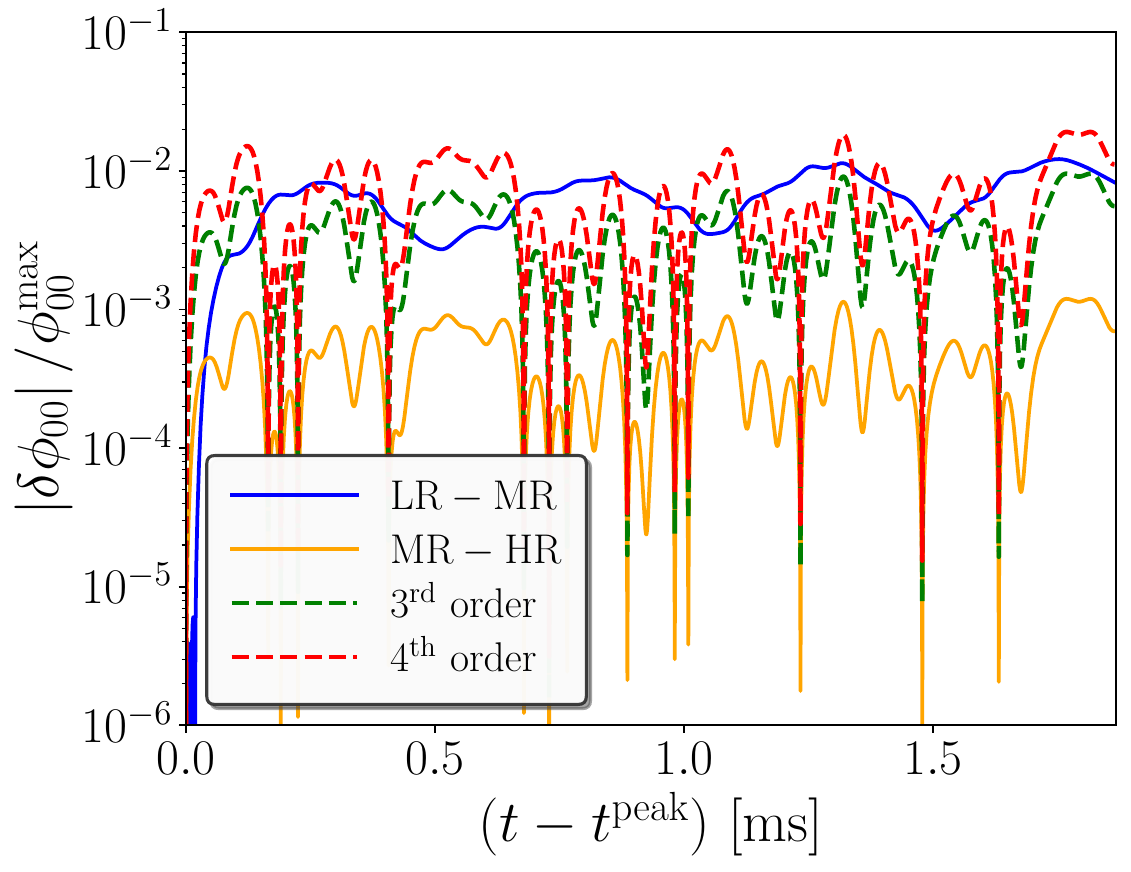}
	\caption{\emph{Convergence test of a single NS.} We show the difference across resolutions of the $(0,0)$ mode of the scalar field for the spontaneous scalarisation of a spinning NS. The difference between the high and medium resolution runs has been rescaled by the convergence factor $Q_n=\frac{h^n_{\rm LR}-h^n_{\rm MR}}{h^n_{\rm MR}-h^n_{\rm HR}}$, assuming third and fourth order convergence.}
   \label{fig:conv}
  \end{figure}

Fig.~\ref{fig:conv} displays the difference across resolutions of the $(0,0)$ mode of the scalar field aligned at the time when the first peak occurs (corresponding to the maximum) and normalised with its value in the peak. It shows that the convergence order is approximately between $3^{\rm rd}$ and $4^{\rm th}$ order, with some mild over-convergence very likely due to the oscillatory nature of the observable we have considered. This is perfectly consistent with the order of the numerical schemes used. We have also verified that both the peak amplitude and the time required to reach it exhibit an order of convergence between $2.5$ and $3$, as expected due to the non-linearities arising during scalarisation.

\

\begin{figure}[h!]
	\centering
     \includegraphics[width=1.0\linewidth]{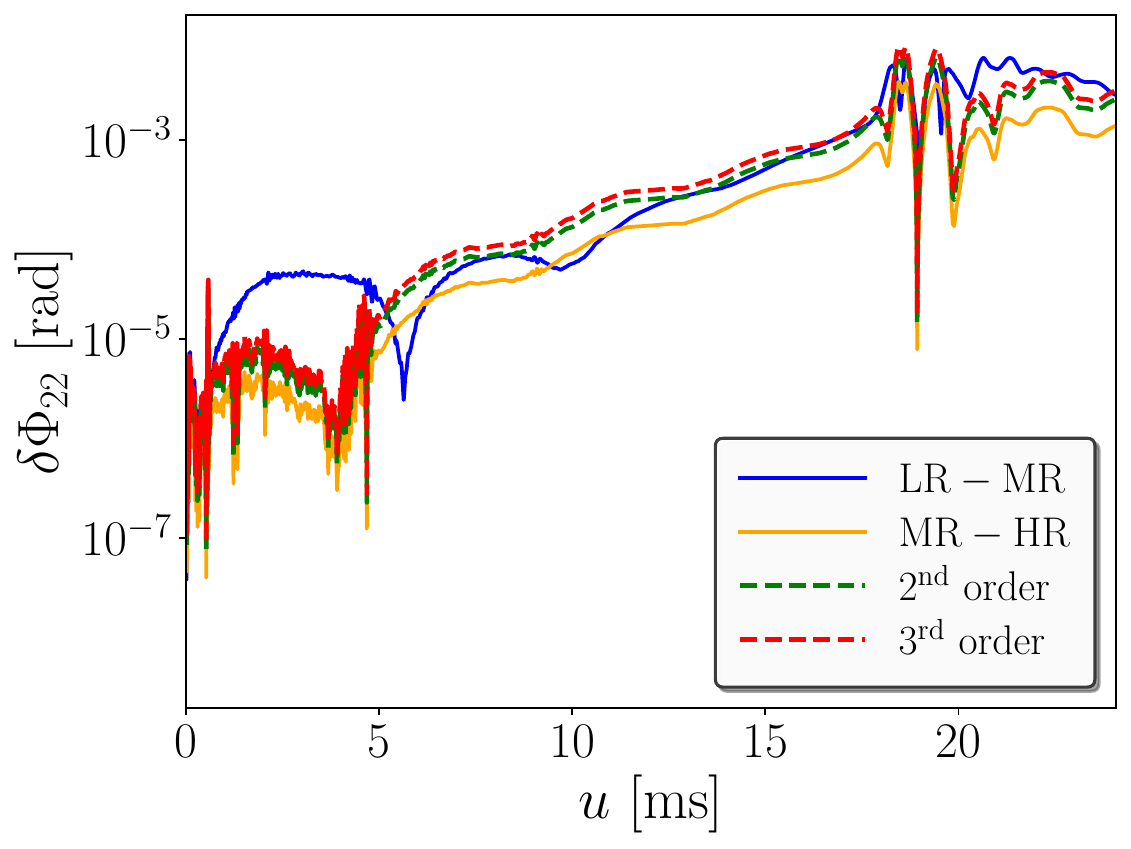}
	\caption{\emph{Convergence test of a BNS merger.} We show the difference across resolutions of the phase of the $(2,2)$ mode of the Weyl-$4$ scalar for a BNS merger yielding spontaneous scalarisation of the HMNS remnant.}
   \label{fig:conv2}
  \end{figure}

{\em \textbf{BNS merger.}---}
Next, we look at the phase of the gravitational $(2,2)$ mode of the Weyl-$4$ scalar for the BNS merger corresponding to the left of Fig.~\ref{fig:spont_bns} and we display in FIg.~\ref{fig:conv2} its difference across resolutions during inspiral, merger and post-merger phases. The three resolutions employed here correspond to $\Delta x_{\rm min} = 150,~120$ and $100$ m. We observe $2^{\rm nd}$ order convergence during most of the inspiral and close to $3^{\rm rd}$ order in the merger phase. This is also perfectly compatible with the order of the numerical schemes used. 

  \bibliography{references}

\end{document}